\documentclass[preprintnumbers,amsmath,amssymb]{revtex4}

\usepackage{amsmath}    
\usepackage{graphicx}   
\usepackage{color}

\newtheorem{mydef}{Definition}

\begin{document}

\title{God may not play dice, but human observers surely do}
\author{Massimiliano Sassoli de Bianchi}
\affiliation{Laboratorio di Autoricerca di Base, 6914 Carona, Switzerland}\date{\today}
\email{autoricerca@gmail.com}   

\begin{abstract}

We investigate indeterminism in physical observations. For this, we introduce a distinction between genuinely indeterministic (\emph{creation-1} and \emph{discovery-1}) observational processes, and fully deterministic (\emph{creation-2} and \emph{discovery-2}) observational processes, which we analyze by drawing a parallel between the \emph{localization} properties of \emph{microscopic} entities, like electrons, and the \emph{lateralization} properties of \emph{macroscopic} entities, like simple elastic bands. We show that by removing the randomness incorporated in certain of our observational processes, acquiring over them a better control, we also alter these processes in such a radical way that in the end they do not correspond anymore to the observation of the same property. We thus conclude that a certain amount of indeterminism must be accepted and welcomed in our physical observations, as we cannot get rid of it without also diminishing our discriminative power. We also provide in our analysis some elements of clarification regarding the \emph{non-spatial} nature of microscopic entities, which we illustrate by using an analogy with the process of \emph{objectification} of human concepts. Finally, the important notion of \emph{relational properties} is properly defined, and the role played by indeterminism in their characterization clarified.

\keywords{Observation \and Quantum measurement \and Creation \and Discovery \and Quantum probabilities \and Localization \and  Non-spatiality \and Control \and Concepts \and Relational properties}

\end{abstract}

\maketitle

\section{Introduction}
\label{Introduction}

According to so-called \emph{creation-discovery view} of reality~\cite{Aerts3,Aerts4}, our observations (also to be understood as measurements, tests, experiments, experiences, etc.) always involve a double aspect: an aspect of \emph{discovery}, through which we obtain information about what is already present in the system under consideration, prior to our observation, and an aspect of \emph{creation}, through which we literally create (or destroy) what is being observed, by means of the observational process itself.

When the first aspect predominates, we are typically in a classical, deterministic, regime, and when the second aspect predominates, we are typically in a quantum, indeterministic, regime. On the other hand, when both aspects are simultaneously present, we are in a mixed, intermediary regime, neither purely classical nor purely quantum, but quantum-like, which cannot be described by phase space or Hilbert space structures.

Within the general conceptual framework of creation-discovery view, a more definite approach, called \emph{hidden-measurement approach}~\cite{Aerts4,Aerts7}, has also been introduced, to address the more specific issue of quantum measurement and the origin of non-Kolmogorovian quantum probabilities. According to this approach, when an experimenter observes a given property of a quantum entity, like, say, the position of an electron (by means of a suitable measuring apparatus), the process can be described as a mechanism of selection of a specific \emph{deterministic} interaction, amongst a collection of possible ones. And since this selection occurs in a way which is beyond the possibility of practical control by the experimenter, the outcome of the observational process cannot be known in advance, but only described in probabilistic terms.

One can show that this very natural hidden-measurement mechanism (which has been explicitely incorporated in a number of macroscopic machine-models~\cite{Aerts3,Aerts4,Aerts13,Aerts2,Aerts5,Massimiliano2}, able to imitate the behavior of microscopic entities) is a possible key ingredient for the explanation of the non-classical structure of the quantum probability calculus. And in that sense, it certainly represents one of today's best candidates for a realistic solution of the longstanding, and still controversial, measurement problem in quantum mechanics.

An interesting consequence of the hidden-measurement approach is that since quantum probabilities would also be related to a situation of lack of knowledge (not about the state of the system, but about the specific interaction which is selected during a measurement), similarly to classical probabilities their nature would be \emph{epistemic}, and not \emph{ontic}. Also, considering that, without losing generality, each individual hidden interaction can be taken to be a deterministic process, quantum probabilities, and therefore quantum mechanics, wouldn't in principle be incompatible with the view of a completely deterministic world as a whole.

The purpose of the present paper is to analyze some further these questions and ideas, and show that even though, in accordance with Aerts' hidden-measurement approach, one can certainly defend that quantum probabilities are epistemic, as they are describable in terms of the experimenter's lack of knowledge about the specific deterministic hidden interaction which is actually selected, nevertheless one cannot conclude that such a situation is compatible with the view of a radical determinism. The reason for this is that, as we will see, the lack of knowledge subtended by quantum probabilities is in fact ineliminable, and therefore, despite their epistemic character, non-classical systems are nonetheless to be considered as genuinely indeterministic. 

To reach this conclusion, we start by recalling, in Section~\ref{Creation and discovery}, a certain number of basic physical concepts. In doing so, we  also introduce the important conceptual distinction between \emph{creation-1} and \emph{creation-2} observational processes (as well as between \emph{discovery-1} and \emph{discovery-2} processes), which as far as we know has not been clearly made in the literature. In Section~\ref{Quantum elastic bands}, we exemplify these notions on a very simple macroscopic physical entity -- an ordinary elastic band, -- by defining and analyzing what we have called the \emph{left-handedness} property. 

Then, in Section~\ref{Genuine indeterminism}, we show that the indeterminism which is inherent in the observation of the left-handedness of an elastic band, due to the observer's lack of control over the experimental procedure, cannot be eliminated, and thus represents a genuine form of indeterminism. In Section~\ref{Genuine non-spatiality}, we use the analysis of the left-handedness' property of the elastic bands to draw a parallel with the localization property of microscopic entities, like electrons, thanks to which we clarify the genuinely \emph{non-spatial} nature of microscopic entities. 

In Section~\ref{Mental processes}, we provide an additional illustration of the fundamental difference between creation-1 and creation-2 observations, by considering the process of objectification (and therefore spatialization) of an abstract concept into a concrete object, operated by a human mind. In Section~\ref{Lack of control}, we explain in which sense indeterminism has to be preserved in certain of our observations, not to reduce our discriminative power, and in Section~\ref{Relational properties} we analyze how indeterminism is linked to the relational character of typical quantum or quantum-like properties. Finally, in Section~\ref{Conclusion}, we offer a concluding comment.

\section{Creation and discovery}
\label{Creation and discovery}

Let us start by recalling what a \emph{property} is. Generally speaking, it is ``something'' a physical entity is able to possess, independently of the type of context it is confronted with. Properties can either be \emph{actual} or \emph{potential}. If they are actual, it means that the successful outcome of the \emph{observational processes} which are used to define them, can always be predicted in advance, with certainty, at least in principle. On the other hand, if they are potential, it means that such successful outcome cannot be predicted in advance with certainty, not even in principle. 

When we say that an entity has a given property, what we generally mean is that the entity possesses it in actual terms. Thus, if the property is potential in a given moment, it simply means it is not possessed by the entity in that moment. However, a potential property is not an impossible property. Indeed, the notion of potentiality expresses the fact that although an entity may not possess a given property in a given instant, it may nevertheless possess it in a subsequent instant, at least in principle. In other words, a property is potential for a given entity if the entity is available in possibly actualizing it, in specific circumstances. 

\emph{Observations}, as understood in the present paper, are processes that can be used to operationally define the properties we associate to the various physical entities~\cite{Massimiliano-O}. They are also called \emph{experimental projects, tests, experimental questions, measurements}, etc., and correspond to practical procedures whose outcomes lead to well-defined ``yes-no'' alternatives.~\cite{Piron1, Piron2, Piron3,Aerts1} A properly defined observation requires the specification of a measuring apparatus to be used, the actions to be performed, and the rule to be applied to unambiguously interpret the results of the process in terms of the (mutually excluding) ``yes'' (the observation is successful) and ``no'' (the observation is unsuccessful) alternatives.

Generally speaking, for a given property there is a number of different observational processes one can use to \emph{equivalently} define and observe the property in question. More precisely, two observational processes are said to be equivalent if, whenever we are in a position to predict in advance the outcome of one of them, we can also predict the outcome of the other, and vice versa. Therefore, in general terms a property is not defined by means of a single observational process, but of an entire \emph{equivalence class} of observational processes. In fact, mathematically speaking, one usually identifies a property with the equivalence class of its observational processes.~\cite{Piron1, Piron2, Piron3,Aerts1}

When all the actual properties of an entity are known, then also its \emph{state} is known, as in general terms the state of an entity corresponds to the set of all its actual properties, i.e., to the collection of all properties that are actual for the entity in a given moment. Of course, since with time some actual properties become potential, whereas some other potential properties become actual, this means that the state of an entity, in general, changes (i.e., evolves) as time passes by. However, not all properties of an entity necessarily change with time: some of them, usually called \emph{intrinsic properties}, or \emph{attributes}, are more stable, and are typically used to characterize the entity's specific identity.

Considering that the state of an entity is, by definition, the collection of all properties that are actual for that entity in a given moment, it's clear that once we fully know its state, we also know all it can be said \emph{with certainty}, in that moment, about the entity. This may erroneously lead one to believe that, accordingly, the outcome of whatever observation we can perform on the entity is in principle predictable with certainty. This belief is usually referred to as the \emph{classical prejudice}, and it has been put seriously into question by the advent of quantum mechanics~\cite{Piron1,Piron2,Piron3}. However, as we will show in the present paper, one can conclude about the groundlessness of the classical prejudice in more general terms, independently of the specificities of the quantum formalism, and of its possible interpretations.

If the outcome of an observation associated to a given property is always predictable in advance with certainty, regardless of the state in which the entity is, then the property is called \emph{classical}, in relation to that entity. A same property can however be classical for an entity, but non-classical for another entity. For example, being localized in a given region of space is a classical property for a macroscopic entity, but a non-classical (quantum) property for a microscopic entity, like an electron.

Conversely, if there are states of an entity such that the outcome of an observation is not predictable with certainty in advance, not even in principle, then the property associated to that observation is called \emph{non-classical}, in relation to that entity.  Quantum properties are of course a typical example of non-classical properties, but not all non-classical properties are necessarily of the quantum kind (i.e., describable by a projection operator acting on a Hilbert space). 

Based on the above definition of a classical property, one could be tempted to define a ``classical entity'' as an entity whose properties would be all classical. However, strictly speaking, classical entities do not exist. Indeed, it is always possible to submit a hypothetical classical entity to an observational process defined in such a way that its outcome will not be predictable in advance, so that the property associated to that observation would be, by definition, a non-classical property of the entity, thus contradicting its presumed classicality. This means that an entity can be considered classical only provided we restrict our observations to a limited subset of classical properties. (See~\cite{Aerts12} for a more elaborated discussion about the influence of experimental contexts in the characterization of the classicality of an entity).

Seemingly, one may be tempted to define a pure ``non-classical entity'' as an entity whose properties are all non-classical. But again, strictly speaking, pure non-classical entities do not exist, as to give a proper meaning to the very notion of an entity one needs to attribute to it, as a minimum, the property of its existence, which by definition is a classical property (at least for the entire duration of existence of the entity). Also, one needs to attach to the entity a given number of attributes, to characterize its distinctive identity, and attributes are by definition classical properties.

As we said, observations can either be of the \emph{discovery} kind or of the \emph{creation} kind. A discovery is a \emph{non-invasive} observational process, which doesn't alter the state of the entity under examination. A creation (in the sense understood in this paper) is an invasive observational process producing a change in the state of the system under examination. In this work we will only consider \emph{soft} creations, i.e., structure preserving processes that do not alter the set of states of the entity. More invasive processes, called \emph{hard} creations, which are able to alter the set of states of an entity and go as far as to destroy its original identity, can also be considered, but we will not need to do so in the present analysis~\cite{Coecke}. 

Observations of the creation kind can be divided into two fundamental categories, which are the following:

\begin{mydef}
\label{definition-1}
\emph{[creation-1 and creation-2 observations]}. A \emph{creation-1} is an invasive observational process whose effects on the observed entity (i.e., the outcomes of the observation) cannot in general (i.e., for all entity's states) be predicted in advance, not even in principle. On the other hand, a \emph{creation-2} is an invasive observational process whose effects are always predictable in advance, with certainty, at least in principle.
\end{mydef}

Similarly, also observations of the discovery kind can be divided into two fundamental categories:

\begin{mydef}
\label{definition-2}
\emph{[discovery-1 and discovery-2 observations]}. A \emph{discovery-1} is a non-invasive observational process whose outcomes cannot in general (i.e., for all entity's states) be predicted in advance, not even in principle. On the other hand, a \emph{discovery-2} is a non-invasive observational process whose outcomes are always predictable in advance, with certainty, at least in principle.
\end{mydef}

Considering the above definitions, we are now in a position to propose a slightly more specific characterization of the notion of classical property, which is the following:

\begin{mydef}
\label{definition-3} 
\emph{[classical and non-classical properties]}. A property is \emph{classical} (in relation to a given entity) if and only if all the equivalent observational processes which operationally define it, are either of the discovery-2 or of the creation-2 kind. On the other hand, a property is \emph{non-classical} if and only if all the equivalent observational processes which operationally define it are of the discovery-1 or creation-1 kind.
\end{mydef}

As we already remarked, two observational processes are said to be \emph{equivalent} if and only if, when we are in a position to predict in advance the outcome of one of them, we can also predict the outcome of the other, and vice versa. However, since by definition creation-1 and discovery-1 observational processes are unpredictable, one cannot use such a criterion to provide a general characterization also of equivalent creation-1, or discovery-1, observational processes. Therefore, we propose to characterize in more general terms the equivalency of two observational processes as follows:

\begin{mydef}
\label{definition-4}
\emph{[equivalent observations]}. Two observational processes are \emph{equivalent} (in relation to a given entity) if and only if they are characterized by exactly the same possible outcomes and \emph{(1)} in case these outcomes would be predictable in advance, the predictions for both observations always coincide, for all entity's states \emph{(deterministic equivalence)}, or \emph{(2)} in case these outcomes wouldn't be predictable in advance, not even in principle, the outcomes' statistics for both observations coincide, for all entity's states \emph{(indeterministic equivalence)}.
\end{mydef}

From such a definition, it immediately follows that creation-1 and creation-2 observational processes (or discovery-1 and discovery-2 observational processes) can never correspond to equivalent observations, and therefore be used to define the same physical property, even though they are characterized by the same outcomes. And this also means that one cannot transform a creation-1 observation into a creation-2 observation (or a discovery-1 observation into a discovery-2 observation) without irremediably affecting the very property which is being observed. It is one of the purposes of the present essay to make this fact as explicit and evident as possible, through the analysis of a simple macroscopic example.

\section{Quantum elastic bands}
\label{Quantum elastic bands}

In this section we will exemplify some of the notions we have introduced in the previous section by studying a very simple physical entity: a (uniform) \emph{elastic band}. Our definition of an elastic band includes the possibility for the elastic to be also broken in multiple fragments. In other terms, any process producing the breaking of the elastic will not be considered as a process that destroys the elastic's identity, but just as a process that changes its state.

It is easy to associate to an elastic band a number of \emph{classical properties}. As an example, consider the property $L_{10}$ of ``having an unstreched length greater than 10 centimeters.'' A possible observational process associated to this property is the following: ``Take a measuring tape, measure the respective lengths of all the elastic fragments, taking care every time to replace them in the exact position they were before the measurement, then add up all the obtained lengths, and if the result is greater than 10 centimeters the observation of the property is successful (outcome `yes'); in the opposite case, the observation is unsuccessful (outcome `no').''

It is of course easy to find other observational processes that can be used to equivalently define such property. For instance, one can utilize a camera and take a picture of the elastic's fragments, placing next to them a reference object, whose length is known, then use the picture to calculate the overall elastic's length. This test is clearly equivalent to the previous one, as it yields exactly the same outcomes. 
Of course, given an arbitrary elastic band entity, if we fully know its state (which also includes a specification of its unstreched length), we can predict in advance, with certainty, the result of the two above mentioned observational tests, which is the reason why $L_{10}$ is (according to our defintion) a classical property. Let us also observe that the equivalent measuring procedures associated to $L_{10}$ are perfectly non-invasive for the elastic-entity, so that they correspond to \emph{discovery-2} processes.   

In a similar way, one can easily define and associate other properties to elastic band entities, by means of (non-invasive) observational processes of the discovery-2 kind, like for instance the property of ``having a volume greater than 0.5 cubic centimeters,'' of ``being made of exactly 3 separate fragments,'' etc. But, as we have emphasized in the previous section, a classical property can also be defined in terms of a creation-2 process. Let us consider for instance the property of \emph{cuttability}, which is so defined: ``Count the number $N$ of fragments forming the elastic band, then take the longest one and cut it in two halves with a scissor, then count again the number $M$ of fragments of the elastic. If $M=N+1$, the observation of the cuttability property is successful (outcome `yes'); otherwise, the observation is unsuccessful and the outcome is the `no' answer.''

Let us highlight a subtle point: the creation-2 observational process associated to the cuttability property is not a process through which the cuttability property is created. It is a process through which the property is simply made manifest. Indeed, the cuttability property was already possessed, in actual terms, by the elastic band entity before its practical observation, as is clear that one could predict with certainty the outcome ``yes'' of the observation before executing it. However, if we execute in practical terms the observation, the process is necessarily invasive, in the sense that it changes the state of the system, here by increasing the number of fragments of the elastic. So, during the observation there is the \emph{creation of a new state}, although \emph{not the creation of the observed property}.

This means that, technically speaking, a creation-2 process is in fact also a discovery process, although of an invasive kind, as it consists in the observation of how a system predictably reacts, by changing its state, when it is acted upon in a specific way.\footnote{This means that we could have made a different terminological choice in this paper. The one we have adopted suggests that a discovery has to be primarily understood as a non-invasive observation, and that a creation can also be understood as a predictable process. The other possibility is to consider that a discovery should be primarily understood as an observation of what is already present in the system, independently of the invasiveness or non-invasiveness of the observational process, whereas a creation should only be such if purely unpredictable.} Let us also notice that the classical character of the cuttability property is related to the fact that its observational process, although invasive, is under the \emph{full control} of the observer, and this is the reason why the outcome is predetermined (i.e., predictable in advance with certainty) and the property classical. Does this mean that an elastic band is a classical entity? As we explained in the previous section, we can only say so if we restrict our observations to a subset of classical properties, as it is also possible to associate non-classical properties to conventional object-entities, like an elastic band. These non-classical properties are defined in terms of creation-1 and discovery-1 observational processes. Let us consider an example of a property defined by means of a creation-1 process, which we will call \emph{left-handedness} (examples of a discovery-1 observational processes will be given in Sections~\ref{Third objection} and \ref{Relational properties}).

\begin{figure}[!ht]
\centering
\includegraphics[scale =1]{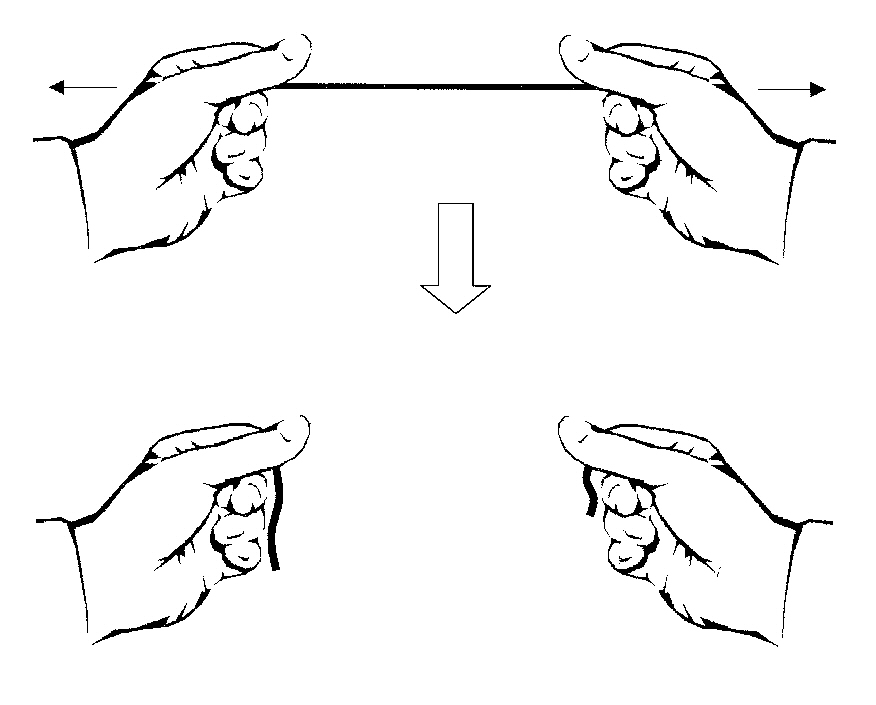}
\caption{A succesful observation of the left-handedness of an elastic band.
\label{lefthandedness}}
\end{figure}

To know what the left-handedness of an elastic band is, all we need to do is to explain how we have to proceed to unambiguously observe it, in practical terms. The observational procedure is the following:

\begin{mydef}
\label{definition-5}
\emph{[left-handedness of an elastic band]}: Grab the two ends of the longest fragment of the elastic, with both hands, then stretch it \emph{strongly and abruptly}, so it breaks. If the longest fragment remains in your left hand, the left-handedness has been successfully observed (outcome ``yes''); otherwise, the observation has been unsuccessful and the outcome is ``no'' (see Fig.~\ref{lefthandedness}).
\end{mydef}

The observation of the left-handedness of an elastic band is a typical example of a process during which a specific interaction is selected among a number of possible ones, in a non predictable way, in accordance with Aerts' hidden-measurement description. Indeed, when the observer stretches \emph{abruptly} the elastic band, s/he has no ways whatsoever to control the final effect of its action, i.e., the point $x$ at which the elastic will eventually break, thus producing either the ``yes'' or the ``no'' outcome. 

One can describe this process by saying that to every possible breaking point $x$, it corresponds a specific interaction $I_x$ (in fact, an entire class $[I_x]$ of equivalent interactions) between the measuring apparatus constituted by the two hands of the observer and the system constituted by the elastic band entity. Because of the presence of a number of fluctuating factors, such as the imperceptible oscillations of the hands, the variable pressure exerted by the fingers, the effective rapidity with which the elastic is abruptly stretched, and so on, it is truly impossible for the experimenter to control which exact interaction $I_x$ will be actually selected, among the countless a priori possible ones, and thus know in advance the point $x$ at which the elastic will finally break (let us recall that the human observer must stretch the elastic strongly and abruptly, to comply with the observational procedure).

Before the execution of the observational experiment, all the breaking points are \emph{a priori} possible: they are all \emph{potential} breaking points. But when the elastic is grasped and stretched, a symmetry breaking process occurs, in the sense that a specific interaction $I_x$ is unconsciously selected by the observer, causing the rupture of the elastic in a specific point $x$. Accordingly, the property of the left-handedness, which was only potential prior to the experiment, will be either actualized or not, i.e., its observation will be either successful or not.

Let us observe that although the outcome of the left-handedness observation is unpredictable in advance, the observer can nevertheless calculate a probability for it. The observational procedure being perfectly symmetric with respect to the left and right hands of the experimenter, the probability to observe the left-handedness is obviously equal to $0.5$ (a fact which can be confirmed by performing a sufficient number of observations on identical elastic bands). This probability, however, doesn't result from a lack of knowledge concerning the state of the observed entity (as it is assumed in certain hidden variables theories), but about the interaction which effectively takes place between the observer's measuring apparatus (her/his hands) and the observed entity.

In other terms, we must distinguish two different categories of lack of knowledge, producing different probability calculus. Classical probabilities, obeying Kolmogorov's axioms, correspond to situations where the lack of knowledge is only about the state of the entity, whereas non-classical probabilities, not obeying Kolmogorov's axioms, correspond to situations where there can be perfect knowledge about the state of the entity, but lack of knowledge about the interaction which is actually selected during the observation (i.e., during the measurement). Quantum probabilities appear then to be a special case of non-classical probabilities, corresponding to the limit situation where the lack of knowledge about the interaction is maximal~\cite{Aerts4,Aerts7,Aerts3,Aerts13}.

Here we would like to point out three important features of the left-handedness observation, which we believe are also typical of a quantum measurement~\cite{Massimiliano-O}: (1) the presence of a random selection mechanism; (2) the irreducible invasiveness of the observational process; and (3) the ephemerality of the observed property, due to its relational character.  

The random selection mechanism is the one which picks, in an unpredictable way, a specific interaction $I_x$, the unpredictability being caused by the fluctuations which are built-in in the very observational procedure. In formal terms, such a random selection mechanism is described in the Geneva-Brussel formalism by the notion of \emph{product test} (or \emph{product observation}, to use the terminology introduced in~\cite{ Massimiliano-O}), originally formulated by Constantin Piron~\cite{Piron1,Piron2,Piron3}. More will be said about product observations in the following.

The irreducible invasiveness consists in the fact that the very observational procedure contemplates the action of breaking the longest elastic fragment into two halves, thus inevitably changing the elastic's state. Such a disturbance cannot be avoided, as the breaking action is encoded in the very operational definition of the left-handedness property.

Finally, the left-handedness property is clearly ephemeral. Indeed, imagine you have just observed, say with success, the left-handedness of a given elastic band, and want to repeat on that same elastic the same observation. According to the procedure, all you have to do is to take again the longest fragment, grab it with your two hands, stretch it abruptly until it breaks and observe if the longest fragment remains once more in your left hand. Clearly, the fact that the previous observation was successful will not help at all in predicting the result of the subsequent one. Indeed, to be re-observed, the left-handedness property, which is a \emph{relational property} between the elastic and the experimental apparatus (formed by the experimenter's hands), has to be re-created, and the creation process is a non-deterministic creation-1 process.

The above three features are most likely also present in a typical quantum measurement. Concerning point (2), it is well known that a quantum measurement produces an inevitable ``wave function collapse,'' thus changing the state of the entity. Also, regarding point (3), we know that most properties of microscopic entities, like for instance the spatial localization of an electron, are ephemeral in nature~\cite{Massimiliano1}. Indeed, quantum mechanics teaches us that, in general, not only can one not predict the localization of a microscopic entity prior to the measurement, but neither a finite time following it, however small such a finite time is. Point (1), being about a ``hidden mechanism,'' obviously it can only be inferred considering that the nature of quantum probabilities are non-Kolmogorovian, and that a random selection mechanism of the hidden-measurement kind can perfectly reproduce their specific non-Kolmogorovian nature~\cite{Aerts7,Aerts10,Aerts11}. (In Section~\ref{Relational properties} we show that point~(3) can in fact be considered as a consequence of points~(1) and (2)).

\section{Genuine indeterminism}
\label{Genuine indeterminism}

We are now in a position to make our point regarding the presence of a genuine form of \emph{indeterminism} in our reality, despite the possible epistemic character of quantum (or quantum-like) probabilities. Here we will understand determinism in the restricted sense of \emph{predictability}, i.e., in the sense that the results of our observations about physical entities are always completely predictable in principle. Thus, we consider indeterminism as a failure of predictability, i.e., as the existence of observations that cannot be predicted, not even in principle. Our point is that the result of the observation of a property of the left-handedness kind is authentically unpredictable, i.e., that the left-handedness' observational process is \emph{genuinely indeterministic}. To see why it is so, we will proceed by considering what we think are the only three reasonable objections to such a statement, and show that they are all inconsequential.

\subsection{First objection: we can acquire a better control}
\label{First objection}

The first objection is that the reason for the unpredictability of the left-handedness experiment is related to the \emph{lack of control} of the experimenter. Therefore, if one allows the experimenter to acquire a better control over the entire experimental procedure, then there would be no problems in predicting with certainty the outcome of the experiment. Therefore, the indeterminism inherent in the observation of the left-handedness property cannot be considered genuine.

To make this objection more precise, let us assume that point $x=0$ corresponds to the end of the elastic band fragment (of, say, total length $L$) which is held by the left hand, and point $x=L$ to the end of it, which is held by the right hand, so that for $0<x<L/2$ the outcome of the observation is unsuccessful (``no'' answer), whereas for $L/2<x<L$ it is successful (``yes'' answer). Special consideration should be given to point $x=L/2$, which in fact corresponds to a different potential property of the elastic, we may call ``ambidexterity.'' However, to avoid unnecessary complications, we assume in our discussion that it is always possible in practical terms to determine which of the two obtained fragments is the longest one (in other terms, we assume that the value $x=L/2$ never occurs).

Clearly, we can distinguish here two sets of interactions: $S_{\texttt{yes}}=\{I_x\in [I_x] / x\in (L/2,L)\}$ and $S_{\texttt{no}}=\{I_x \in [I_x] / x\in (0, L/2)\}$, where $[I_x]$ denotes the class of equivalent interactions which deterministically produce the same outcome, with the elastic broken at point $x$ and the fragment of length $x$ (resp. $L-x$) in the left hand  (resp. right hand) of the experimenter. Now, if the experimenter can control the experimental procedure in order to determine in advance if the applied interaction will be in $S_{\texttt{yes}}$ or in  $S_{\texttt{no}}$, then no doubts s/he will be in a position to predict with certainty, in advance, the outcome of the left-handedness observation. 

To see how s/he can do this, let us consider a specific deterministic creation-2 process, $\tilde{I}_x$, which is defined as follows: ``Take a ruler and measure the distance $x$ from one of the two ends of the elastic fragment of length $L$, then using a scissor cut the elastic fragment exactly at point $x$. At this point, grab the fragment of length $x$ with your left hand and the remaining fragment of length $L-x$ with your right hand.'' Clearly, $\tilde{I}_x\in [I_x]$. The only difference with respect to the previously defined $I_x$, is that they are randomly selected by the observer, in a way s/he cannot consciously control, whereas $\tilde{I}_x$ can be selected in a perfectly conscious and controlled way. Thus, a perfectly equivalent definition for the left-handedness property would be the following:

\begin{mydef}
\label{definition-6}
\emph{[left-handedness of an elastic band: equivalent definition]}: Take the longest fragment of the elastic and measure its length $L$ with a ruler. After that, use a random number generator (RNG) to select a value $x$ in the interval $(0,L)$, excluding a priori the specific value $x=L/2$. Next, applies on the elastic fragment of length $L$ the deterministic interaction $\tilde{I}_x$, then observe your hands: if the longest fragment is in your left hand, the left-handedness has been successfully observed (outcome ``yes''); otherwise, the observation has been unsuccessful and the outcome is ``no.''
\end{mydef}

This alternative definition is perfectly equivalent to the original one, as is clear from the fact that the procedure not only produces the same unpredictable outcomes, but it does so with exactly the same probability distribution (see Def.~\ref{definition-4}, Sec.~\ref{Creation and discovery}). The advantage of this alternative equivalent definition is that it makes fully manifest what is the part of the procedure which is effectively under the control of the experimenter -- the execution of the $\tilde{I}_x$ interaction -- and the part which cannot be controlled by her/him -- the selection of the $x$ value by the RNG. So, acquiring a better control over the experimental procedure of the left-handedness property would simply mean to replace the RNG selection mechanism by a more determinative process. But the only way to do so is to redefine the left-handedness observational protocol. Let us call our two previous equivalent definitions \emph{left-handedness-1}, to distinguish them from the property of \emph{left-handedness-2}, which we are now going to define.

\begin{mydef}
\label{definition-7}
\emph{[left-handedness-2 of an elastic band]}: Take the longest fragment of the elastic and measure its length $L$ with a ruler. After that, use a RNG to select a value $x$ in the interval $(0,L/2)$. Next, applies on the elastic fragment of length $L$ the interaction $\tilde{I}_x$, then observe your hands: if the longest fragment is in your left hand, the left-handedness-2 has been successfully observed (outcome ``yes''); otherwise, the observation has been unsuccessful and the outcome is ``no.''
\end{mydef}

Contrary to left-handedness-1, which is a creation-1 process, left-handedness-2 is evidently a creation-2 process, now perfectly deterministic. In other terms, by allowing the experimenter to acquire a full control over the experimental procedure (relative to its outcomes), we have succeeded in eliminating the source of unpredictability which was present in the observational procedure associated to left-handedness-1, thus obtaining a fully deterministic process. This because the outcome ``yes'' is now certain in advance, and left-handedness-2 is an actual property of every elastic band. But as the saying goes: ``operation successful patient dead!'' Indeed, according to Definition~4, left-handedness-1 and left-handedness-2 cannot any more be considered as equivalent observational processes, and therefore \emph{they do not define anymore the same property}. This means that if we increase the level of control of the experimenter, much enough to eliminate the outcomes' unpredictability, the result is that s/he will not observe any more the same property, and this is the reason why left-handedness-1 is a genuine creation-1 process, whose built-in indeterminism cannot be eliminated.

\subsection{Second objection: non-ordinary properties are not bona-fide properties}
\label{Second objection}

According to the above discussion, it is quite evident that one cannot get rid of the unpredictability inherent in the non-classical left-handedness-1 property, as to do so one has to modify the property's operational definition in such a radical way that the modified measurement does not correspond anymore to the specification of the same property, but of a radically different new property, which we have called left-handedness-2. In fact, all this is quite obvious (but obviousness is what needs to be said, to make it obvious), as there are no logical reasons to consider an indeterministic creation-1 process observationally equivalent to a deterministic creation-2 process, as we already emphasized at the end of Sec.~\ref{Creation and discovery}. What we have tried to do is to make this point as clear and explicit as possible, using a very simple and concrete example, where every operation is under our eyes. 

Of course, the objection then is that the way we have defined the left-handedness-1 property (either in terms of the abrupt stretching procedure, or of the equivalent scissor-RNG procedure) is such that we have included into it, \emph{ad hoc}, a condition of lack of control (and therefore an element of randomness), so that, as in a circular reasoning, indeterminism has been proved by inserting on purpose indeterminism in the very observational procedure. This brings us to the second possible criticism: that properties like the left-handedness-1 are not to be considered as bona-fide properties of physical entities, and therefore should not be used to establish the existence of an irreducible indeterminism in our physical reality. 

Undoubtedly, indeterminism is a built-in element in the definition of the left-hadedness-1, but, as we will now see, there is no logical fallacy in this. First of all, let us recall, as we previously mentioned, that observational procedures involving an unpredictable selection mechanism are generally called \emph{product tests} (or product observations), and are used to opportunely define \emph{meet properties}, i.e., properties made by the conjunction of other properties~\cite{Piron1,Aerts11,Aerts12}. Meet properties express a self-evident fact of reality: that physical entities can possess (in potential or actual terms), more than a single property at once, and product tests are just the right procedures we must employ to operationally define and test meet properties. Obviously, when two properties are compatible, we can easily observe them (i.e., test them) simultaneously. For instance, a car can have the properties of being red colored and longer than 3 meters, and there are no problems in conjunctly testing these two properties at the same time. But not all properties are observationally compatible. A piece of wood for instance, possesses both the burnability and floatability properties, and therefore also possesses the meet property of ``burnability \emph{and} floatability.''~\cite{Aerts1}. However, if we burn the piece of wood to test burnability, we will automatically destroy floatability, and vice versa, if we bathe the piece of wood to test floatability, we will destroy (although only temporarily) burnability. In other terms, the observational tests associated to these two properties are mutually incompatible, and therefore cannot be performed together, at the same time. 

But, as shown by Piron, in ultimate analysis there is no need to do so. Indeed, all we need to do is to perform a \emph{product observation} consisting in first selecting randomly (i.e., in a way that cannot be predicted by the observer) either the burnability or the floatability test, and after that performing it. Then, if the outcome is successful (unsuccessful), we will say that also the (product) observational test of the meet property is successful (unsuccessful), i.e., that we succeeded (failed) in observing the ``burnability \emph{and} floatability'' meet property.

To understand why the above procedure is meaningful, we must keep in mind that an entity does actually possess a property if and only if it is possible to predict in principle, with certainty, that its observation would be successful, should we decide to perform it~\cite{Piron1,Piron2,Piron3}. The key point here is that we don't need to perform in practice an observational test to infer the actuality of a property: it is sufficient to be in a position to predict (in principle) \emph{with certainty} its successful outcome. Now, seeing that the product test procedure involves a non-deterministic selection mechanism, the only way we can guarantee that the observational process would be successful, should we perform it, is to know in advance that its success is independent of such a non predictable choice. And this can only be the case if the two properties -- the burnability and floatability ones, in our example -- are both actual. 

So, if we admit that meet properties are testable properties (i.e., definable in operational terms and therefore observable in practice), and considering that product tests are the right procedures to be employed to do so (at least in relation to meet properties made of mutually incompatible properties), we have then to admit that we are not always in a position to predict with certainty the behavior of physical entities, as we cannot do it when we observe certain of their meet properties (see also the discussion in~\cite{Massimiliano-O}). And this would mean that a certain level of indeterminism must be accepted in our description of reality, being an unavoidable structural ingredient of our observational processes.

But of course, one could also defend the point of view that not all the properties we can theoretically associate to a physical entity have necessarily to be observable/testable in practical terms. In other terms, one could consider that meet properties are a typical example of concepts which are maybe useful in theoretical terms, but not for this necessarily always directly testable. In conventional quantum mechanics for instance, we know that a number of important properties are not testable, i.e., associable to a closed linear subspace of the Hilbert space~\cite{Smets}. For instance, ``being in a state $\psi$ or in a state $\phi$'' is not a testable property of a quantum entity, i.e., it is not a property that can be associated to a projection operator (unless the two states are orthogonal). This is generally the case also for meet properties made by the assemblage of properties whose associated projection operators are non-commuting. For instance, for a microscopic entity like an electron, the meet property $a(R)\wedge b(C)$ [to be read: ``$a(R)$ \emph{and} $b(C)$''], where $a(R)$ is the property of being localized in a spatial region $R$, and $b(C)$ is the property of having the momentum lying in a cone $C$,'' cannot be associated to a projection operator, since the projection operator $P_R$ associated to $a(R)$, and the projection operator $P_C$ associated to $b(C)$, do not commute. Accordingly, one could be tempted to conclude that the meet property $a(R)\wedge b(C)$ is not a testable (i.e., directly observable) property.   

On the other hand, the previously mentioned product test (or product observation) protocol, tells us that also meet properties of the form $a(R)\wedge b(C)$ are in fact observable, and this independently of the fact that they are made of experimentally incompatible properties. Indeed, from the ampler perspective of a general operational description of physical systems, meet properties are always in principle observables, by means of product observations. Of course, if we accept this view, we must also accept that conventional quantum mechanics is an incomplete description, as its formalism cannot account for certain observational possibilities. In fact, that quantum mechanics is an incomplete theory, in the sense that serious structural shortcomings would be present in its formalism, is already an established fact, as for instance it has been clearly demonstrated that one cannot accommodate the description of separated physical entities within its too restricted Hilbert space structure~\cite{Aerts1}. However, our purpose is not to discuss here the shortcomings of conventional quantum mechanics, but to understand if properties like the left-handedness-1 are physically meaningful or not, in a general operational description of physical systems. Because if they are, then a certain level of indeterminism cannot be avoided in our description of the world.

To resolve this issue, we need to reflect on how we usually ascribe properties to physical entities. For instance, why do we attach to the moon the intrinsic property of \emph{spatiality}, i.e., the property of always having a specific location in space, in every moment? One possible response is that since in our past interactions with solid macroscopic objects we have always found them in a specific spatial position (although this position can possibly change over time), it is quite natural to assume that also our satellite does always possess one, independently of the fact that we may know it or not, or that we are presently observing it or not. 

Roughly speaking, we can say that we humans, in the course of our evolution, have gained some knowledge (through experience) about the nature and behavior of the macroscopic material entities forming our reality, by \emph{discovering} that a number of intrinsic properties can sensibly be applied to them. And these properties we have in this way associated to these entities are typically what we have called \emph{classical properties}. However, from such experience, whether we may like it or not, a \emph{prejudice} resulted: the belief that we could generalize, without problems, the results of our observations. For instance, by considering that since the concept of spatiality can suitably be applied to all the \emph{ordinary} macroscopic objects of our everyday life, the same must hold true also for those \emph{non-ordinary} microscopic ``objects'' we are able to detect through the more sophisticated instruments of our laboratories, which extend the observational power of our ordinary senses.

But we must be careful not to commit what logicians call the fallacy of \emph{hasty generalization}, reaching a general conclusion on the basis of information obtained on a sample which is not necessarily representative. Obviously, there is no logical basis for such an inference, as we have never directly interacted, through our ordinary senses, with an individual microscopic entity. The aforementioned prejudice, as we will see, is at the basis of our lack of understanding about the true nature of microscopic entities, as in fact it is a false prejudice. What we would like to highlight here is that: \emph{it is a non-ordinary procedure to attribute ordinary (classical) properties to non-ordinary entities, such as electrons, in the same way  as it is a non-ordinary procedure (although for opposite reasons) to attribute non-ordinary (non-classical) properties to ordinary entities, such as elastic bands}.
 
What we are here emphasizing, as we anticipated in Section~\ref{Creation and discovery}, is that what determines the classical or quantum nature of certain physical entities is not the fact that they are microscopic or macroscopic, but the nature of the properties we believe we can meaningfully associate to them, and therefore observe in practical terms. The important point to understand is that the classical or non-classical character of a property does not depend on the property in itself, but on its specific relation with the entity to which the property is associated, and possibly observed (i.e., tested). If the spatial localization of the moon can be associated to an ordinary observational procedure of the discovery-2 kind, the same we cannot say about the spatial localization of an electron.

So, if one defends the view that properties like the left-handedness-1 cannot to be considered as bona-fide properties of physical entities, because they would be associated to non-ordinary procedures, then the same must be said, \emph{mutatis mutandis}, for the property of spatial localization associated to a microscopic entity like an electron. Thus, the conclusion is: either we consider that only pure discovery procedures, associated to ordinary observational processes, are allowed in our description of reality (and this, by the way, would also be a quite radical position to adopt), and then we have to stop speaking about the observation of the localization of an electron, but also of its momentum, energy, spin, etc.; or else, we accept that our observational processes necessarily extend beyond mere discovery procedures, and then creation-1 processes must also be considered as bona-fide observational processes.

\subsection{Third objection: we can observe the observer}
\label{Third objection}

We come now to the third objection, which from a certain perspective may be considered as the more serious one. Let us call the observer performing the observation of the left-handedness-1 property observer-A. As we previously explained, because of the impossibility of acquiring a better control over the observational procedure without altering it, observer-A is not in a position to predict the result of her/his observation, which therefore has to be considered, at least for her/him, as genuinely indeterministic. 

But what about a second observer -- let us call it observer-B -- observing the scene. Let us assume that observer-B has gathered enough knowledge about the state of observer-A and the state the elastic band, and that thanks to this knowledge s/he is in a position to fully describe the deterministic evolution of the double-system ``elastic-observer-A'' in the laboratory context, and therefore also predict its future states with certainty. 

In other words, if observer-B has sufficient computing power to simulate every detail of the evolution of the ``elastic-observer-A'' composite system in the laboratory context, taking into account the tiniest possible fluctuations, then s/he will be in a position to determine in advance, at least in principle, what will be the outcome of observer-A's experiment with the elastic, i.e., if the longest fragment will dangle from her/his left or right hand.

Let us assume, in accordance with the deterministic hypothesis, that observer-B can actually predict with certainty that the outcome of observer-A's observation of the left-handedness-1 property will be successful, i.e., that the final state of the elastic band will be a left-handed state. Accordingly, one may be tempted to conclude that the observation of the left-handedness-1 property, while an indeterministic process for observer-A, is a deterministic process for observer-B. Consequently, the indeterminism subtended by the observation of the left-handedness-1 property by observer-A would not be genuine, but just an illusion, as the outcome of the observational process appears to be predictable by observer-B. 

But such a conclusion is not permitted. Indeed, observer-B is not actually observing the same entity as observer-A, nor the same property. Observer-A is observing the left-handedness-1 property of an elastic band, and to do so s/he has to apply a specific invasive procedure, which operationally defines the property in question. The only way observer-A has to predict the result in advance, is to increase her/his control, but as we already discussed in Sec.~\ref{First objection}, if s/he does so s/he will not any more observe the same property. Therefore, for observer-A the indeterminism inherent in the observational process is irreducible. 

On the other hand, the observational process of observer-B is completely different: it is apparently a non-invasive procedure referring to a bigger system, made of the elastic band \emph{plus} observer-A. It is this bigger ``elastic-observer-A'' double-system that observer-B is actually observing, and the observational protocol is the following: ``Let the system, \emph{in a given known initial state}, evolve by itself and, by sight, just look at the components called `human hands' until they have broken the component called `elastic band'; if the subcomponents thus obtained, called `fragments of the elastic band,' are such that the longest one is associated with the left hand, then the observation is considered successful; otherwise, it is considered unsuccessful.'' 

In other terms, assuming that observer-B knows the initial state of the double-system and is able in principle to predict its evolution, its observational process is of the \emph{discovery-2} kind. Also, its observation is not about a property of the single elastic band, but of the ampler ``elastic-observer-A'' double-system. Of course, it may be objected that even though observer-B is observing a bigger system, by doing so s/he is also observing its components, so that observer-B is also observing the elastic band, and therefore, indirectly, s/he is observing its left-handedness-1 property. 

This however would be again an illicit conclusion. Indeed, even if we reinterpret the above experimental protocol as an observational protocol only referring to the elastic band, clearly such a protocol cannot be considered as equivalent to the one associated to the left-handedness-1 observational process, described in Sec.~\ref{Quantum elastic bands} (Def.~\ref{definition-5}), considering our definition of equivalency, given in Sec.~\ref{Creation and discovery} (Def.~\ref{definition-4}). So, even considering the observation of observer-B as an observation of the elastic band, we must conclude s/he is not observing the same property of it, so that the fact s/he can in principle predict with certainty the outcome of the observation is not in contradiction with the presence of a genuine form of indeterminism inherent in observational processes of the left-handedness-1 kind. 

Also, in the above description of the discovery-2 observational procedure of observer-B, it wasn't specified by which means the initial state of the double-system happened to be known. Actually, to know the initial state one needs to \emph{prepare} the system in that state, and if we include, as we should, such a preparation procedure into the operational protocol of observer-B, then we cannot anymore affirm  that its observation would be of the discovery-2 kind, but in fact of the \emph{creation-2} kind. But then, also in this case, the same reasoning as per above applies. On the other hand, in case observer-B wouldn't prepare the double-system in a given, known state, then s/he wouldn't be in a position to predict the outcome of the observation, and therefore her/his observational process would be an indeterministic \emph{discovery-1} process. Thus, we must conclude that a genuine, irreducible form of indeterminism has to be accepted as an integral part of our description of reality.

\section{Genuine non-spatiality}
\label{Genuine non-spatiality}

In this section we will strengthen our point regarding the genuine indeterminism built-in in our observation of the world, by drawing a suggestive parallel between the left-handedness observation on a macroscopic elastic-entity, and a spatial localization observation (measurement) on a microscopic entity. Before doing so, let us just emphasize that, according to the logic of our previous discussion, a powerful strategy to understand the puzzling nature of microscopic entities is clearly to ask non-ordinary operational questions in relation to ordinary macroscopic entities, and see how they respond to these questions in practical terms. This is the program that was initiated by Aerts many years ago, when he conceived a number of mechanical macroscopic machines able to imitate the quantum behavior of microscopic entities~\cite{Aerts3,Aerts4,Aerts13,Aerts2,Aerts5,Massimiliano2}, revealing in this way the hidden measurement mechanism that is possibly at the origin of quantum probabilities~\cite{Aerts4,Aerts7}.

Having said this, to draw a parallel between the non-classical left-handedness observation and a typical localization observation, let us consider the property of ``being localized in $R$,'' with $R$ a given arbitrary region of space (not necessarily bounded). Let us call it the \emph{$R$-localization} property. For a classical point particle, \emph{$R$-localization}  is obviously a classical property, but for a quantum microscopic entity it is a typical non-classical property. Indeed, if $\psi_t$ is the state of the entity at time $t$, and $P_R$ the projection operator onto the set of states localized in the region $R$, then the probability $\| P_R \psi_t\|^2$ of finding the entity inside $R$, at time $t$, is in general different from $1$ or $0$. This means that the two possible outcomes of the observation -- ``yes,'' the entity is present inside $R$, or ``no,'' the entity is not present inside of $R$ -- are not predetermined.

In other terms, denoting by $\bar{R}=\mathbb{R}^3\backslash R$ the complement of $R$ in $\mathbb{R}^3$, we can say that, before the observation, the microscopic entity, say an electron, is in a \emph{superposition state} $\psi_t = P_R \psi_t + P_{\bar{R}}\psi_t$ with respect to the alternative of being present in one of the two mutually excluding regions $R$ and $\bar{R}$, in the sense that neither the $R$-localization, nor the $\bar{R}$-localization, are actual properties of the electron. Similarly, if we define the \emph{right-handedness} property by means of the same observational protocol as for the \emph{left-handedness}, but reversing the ``yes-no'' alternative, we can say that also the elastic band, before the observation, is in a superposition state with respect of the two possible outcomes of the experiment, in the sense that neither the left-handedness, nor the right-handedness, are actual properties of the elastic, but just potential properties of it. 

Now, considering this deep analogy between the electron's localization measurement, and the elastic band's lateralization observations (in both cases the outcome is only predictable in probabilistic terms, although the state of the system is perfectly known), what can we conclude about the \emph{spatiality} of a microscopic entity? According to the logic of our analysis, we are forced to conclude that neither the $R$-localization, nor the $\bar{R}$-localization of the electron are properties which pre-exist their observation, and therefore, contrary to the ordinary objects of the macroscopic world, an electron \emph{cannot be considered as an entity generally present in our three-dimensional space}. In other words, an electron does not actually possess a specific location, except in the moment it is detected by an instrument of observation, in the same way as an elastic band does not actually possess a specific (type 1) left or right \emph{lateralization}, except when such a lateralization is actualized (created) through an observation. 

This is also so because, among other things, Heisenberg's uncertainty principle forbids us to observe simultaneously the position of the electron and how its position varies locally in time, i.e., its velocity, so that we have no possibility (by solving the classical equations of motion) to predict with certainty the outcome of a localization measurement, and therefore the reality criterion for such a property no longer applies~\cite{Massimiliano1}. There are in fact many other reasons we could advocate to conclude about the surprising \emph{non-spatiality} of microscopic entities, but we will not do it here, not to overly extend an already long article, and simply refer the interested reader to the following papers (and the references cited therein), where this fundamental notion is extensively explored and discussed~\cite{Aerts4,Aerts2,Massimiliano1,Massimiliano2,Massimiliano3} 

It is worth noticing that microscopic entities, such as electrons, while being non-spatial entities, are nevertheless fully \emph{available} to relate with the macroscopic entities forming our ordinary three-dimensional space, in the sense that they are fully available in manifesting their presence in this space, in the ambit of an observational process, i.e., of a possible interaction with a macroscopic entity, playing the role of the measuring apparatus. This means that although there is no way in general to predict with certainty whether the observation of an electron (of which we assume to know the state) in a given region $R$ of space (or in its complement  $\bar{R}=\mathbb{R}^3\backslash R$), will give a positive or negative outcome, we can nevertheless predict with certainty that if the observation extends to the entire three-dimensional space, then the probability of detecting its presence would become equal to 1, as is clear that, because of the normalization of state vectors, $\| (P_R+P_{\bar{R}}) \psi_t\|^2 = \| \psi_t\|^2 =1$, for all $t\in \mathbb{R}$. 

Here we find ourselves in a situation very similar to that of the experiment about the left-handedness-1 (or right-handedness-1) of the elastic band. Indeed, as well as the elastic band is fully available to interact with the two hands of the human observer, in order to actualize either the left-handedness-1 or the right-handedness-1 property, in the same way an electron is fully available to interact with the two spatial regions $R$ and $\bar{R}$ (more precisely, with the position detectors which are placed within them), to actualize either the $R$-localization or the $\bar{R}$-localization property.

These localization properties are literally created at the very moment of their observation, and therefore it wouldn't make sense to say that before the observation the electron was \emph{actually} present in one of the two regions. The only thing we can say is that it was \emph{potentially} present in both regions, in the same way as an elastic band is potentially both left-handed-1 and right-handed-1. We cannot say either that the electron, before the observation, was simultaneously actually present in both regions, as this means that it would be simultaneously detectable in both regions, whereas an electron can only show itself, when observed, in a single spatial location at a time (as well as a left-handedness-1 experiment can only give a positive or negative outcome, but not both outcomes simultaneously).

Also, as we must not confuse the full availability of an elastic band (for the very fact of its existence) to participate in a left-handedness-1 experiment, with the possible outcomes of the experiment, so we must not confuse the full availability of an electron (for the very fact of its existence) to participate in an experiment of spatial localization, with the possible outcomes of the experiment. In fact, while it is true that an electron is always observable in our physical three-dimensional (ordinary) space, that is, detectable in it with probability equal to 1, this doesn't mean that it actually possesses a specific localization inside of it. We have here a typical example of a situation in which the assumption of \emph{reductionism} does not apply. Indeed, while on one hand we can affirm that the electron is \emph{globally present} in the entire three-dimensional physical space (in the sense of being detectable in it with certainty), at all times, on the other hand we cannot conclude from this that it is locally present in one of its regions, i.e., in one of its parts. And this means that in general the nature of the relationship of an electron with the totality of the three-dimensional space cannot be deduced from its relationship with its parts. And of course, we approach here the true mystery of the microscopic world, which turns out to be populated by entities whose non-spatial nature is quite different from that of macroscopic entities, and therefore cannot be described and understood in the same way.

When we are dealing with microscopic entities, such as electrons, the metaphor that consists in thinking of our three dimensional physical space as a \emph{container} no longer applies. Or, rather, it does not apply if we consider the observation of the spatial position of an electron in the usual sense of a process through which the electron would reveal an already acquired presence in a given region of space, and not as a process through which an electron would be \emph{forced} to manifest its presence in that region. 

By this, we mean that because an electron is fully available in \emph{globally} manifesting itself in spatial terms, it is always possible to exploit this global availability to confer to its manifestation a specific, predetermined outcome. We can do this in the same way in which we can exploit the full availability of an elastic band in letting itself be cut, to define a different concept of left-handedness, that we have called left-handedness-2, which corresponds to a property actually possessed by the elastic band, even before its observation (as the successful outcome of its observation can be known in advance).

For example, we can in principle always apply in region $\bar{R}$ a repulsive force field, so that the probability for the electron to be detected in $\bar{R}$ will tend to zero, as the intensity of the repulsive field increases. In other words, by increasing the repulsiveness of the applied field in region $\bar{R}$, it is possible to \emph{increase the control} of the observer over her/his spatial detection experiment. Let us call \emph{$R$-localization-2} the property of an electron to be detected in region $R$, when in the complement region $\bar{R}$ a virtually infinite repulsive field is applied (we limit here our discussion to non-relativistic quantum mechanics, to avoid problems with Klein paradox). Obviously, we can then affirm that the electron actually possesses the \emph{$R$-localization-2}, as is clear that we can predict with certainty that, if we carry out the observational experiment, the electron would inevitably be detected in region $R$. We can therefore say that an electron certainly always possesses, in actual terms, a \emph{type 2 localization}, whereas it doesn't possess a localization in the ordinary sense, which we will call \emph{localization-1}, to distinguish it from localization-2.

\begin{mydef}
\label{definition-8}
\emph{[localization-1 and localization-2]}. An entity possess the \emph{localization-1} property (and is said to belong to the \emph{ordinary physical space}) if, for every (sufficiently large) arbitrary region $R$ ($R\neq\mathbb{R}^3$) of the three-dimensional Euclidean physical space, its detection in either $R$ or in the complement  region $\bar{R}=\mathbb{R}^3\backslash R$, is predictable in advance with certainty (i.e., either $R$-localization-1 or $\bar{R}$-localization-1 is actual for the entity). On the other hand, an entity possesses the \emph{localization-2} property if, for every arbitrary region $R$ ($R\neq\mathbb{R}^3$) of the three-dimensional Euclidean physical space, when the entity is prevented (by some active form of control) to manifest in the complement region $\bar{R}$, it is detectable in $R$ with certainty.
\end{mydef}

The observation of localization-2 is a process that, like the observation of left-handedness-2, is entirely under the control of the experimenter, in the sense that the outcome of the observational process can be predicted with absolute certainty. This means that localization-2 is a property that a microscopic entity, like an electron, always possesses in a stable way, independently of whether it is or is not observed in practical terms, i.e., made manifest. It expresses the full availability of such entity to relate with the structure of the three-dimensional space \emph{as a whole} (and more exactly with the macroscopic entities that characterize it), through its ``confinability'' in a specific region of the same (in the sense of being detectable with certainty in that region), when its presence in any other region is actively prevented. In other words, localization-2 is a weaker form of spatial localization which is generally possessed by microscopic entities, allowing them, under certain conditions of active control, to deterministically relate with specific parts of our space. Now, since to relate to the three-dimensional space and its parts means, in ultimate analysis, to relate with the macroscopic entities that inhabit it, localization-2 is nothing but the property allowing a microscopic entity to \emph{bind} to a macroscopic entity, that is to form what is usually referred to as a \emph{bound state}, so acquiring the same localization-1 property of the latter. 

When a microscopic entity binds to a macroscopic entity, that is, becomes part of it, not only it manifests the property of localization-2, but in fact also acquires (i.e., actualizes) in a sense the localization-1 property possessed by the macroscopic entity. Exactly in the same way as an elastic band, when it manifests its left-handedness-2, obviously also manifests, in the same moment, left-handedness-1, since the successful outcome of the experiment characterizing left-handedness-2 is the same as the one determining left-handedness-1. Thus, we can affirm that microscopic entities are entities that usually do not possess the property of belonging to the ordinary three-dimensional (Euclidean) space, as they do not actually possess, permanently, the localization-1 property, which is possessed instead by macroscopic entities. On the other hand, because they do possess the localization-2 property, which is an expression of their full availability in binding with specific macroscopic entities, they can nevertheless enter and reside permanently in the ordinary physical space, in the form of aggregates, when they are forced to do so under some active form of control. But if a macroscopic aggregate stably possesses the property of localization-1, and therefore can be considered an (ordinary) spatial entity, we must be careful not to think of its individual components as entities that would also possess the same property. Contrary to the hypothesis of reductionism, it is not the spatiality of the microscopic constituents that confer to the macroscopic body its spatiality: spatiality is an emergent property, that the individual components only possess when all together, for as long as they maintain their specific and exclusive relationship, but will immediately lose as soon as their connection from the macroscopic aggregate is severed.

\section{Mental processes and conceptual entities}
\label{Mental processes}

A major difficulty in understanding the emergence of spatiality of macroscopic entities, from the non-spatial layer of microscopic entities, could be our insistence in thinking about the latter in mere \emph{objectual} terms. Still, one may wonder what other picture we would have at our disposal to mentally visualize the entities of the microworld, which in addition to their staggering lack of spatiality present a number of other oddities, like their well-known \emph{lack of distinguishability}, their tendency to produce  \emph{interferences}, form connections of all kinds regardless of spatial distances (\emph{entanglement}), and so on.

A possible answer comes to us once again from the pioneering work of Aerts. Indeed, recently the already broad explanatory framework of the creation-discovery view has been further expanded in order to embrace an even more general class of systems, called \emph{state-context-property} systems (SCoP), thanks to which it becomes possible to describe not only the action of an experimental context on a given system (as is usual in physics), but also the influence of the system on the context itself~\cite{Aerts12}.

The advantage of such a general approach, among other things, is to allow the description not only of physical entities (classical, quantum or quantum-like), but also of more abstract entities, such as human concepts, human minds, and the decision processes associated with them. This has allowed to discover that such a generalization of the formal structure of quantum mechanics lent itself surprisingly well to the construction of a quali-quantitative theory of human concepts and their combinations (see the references cited in~\cite{Aerts-c1,Aerts-c2,Aerts-c3,Aerts-c4}).

Obviously, we cannot enter here into the details of this theory, which is very articulated and would go beyond the scope of the present article. What we want to highlight is just that following the discovery that the structure of the quantum formalism is perfectly able to model human concepts, a rather unusual, but not less natural idea quickly emerged, summarizable in the following question~\cite{Aerts-c1}: ``If quantum mechanics as a formalism models human concepts so well, perhaps this indicates that quantum particles themselves are conceptual entities?''

This fascinating interrogative has given rise to a very innovative interpretation of quantum physics, which is based precisely on the assumption that the nature of quantum entities would be \emph{conceptual}, in the sense that these entities would interact with the macroscopic measuring instruments (and more generally with the entities made of ordinary matter) in a  way similar as how human concepts interact with human minds (or other memory structures sensitive to the meaning of the concepts)~\cite{Aerts-c1,Aerts-c2,Aerts-c3,Aerts-c4}.

As is known, human concepts are typically non-spatial entities. Indeed, one can hardly say that they belong to our three-dimensional space, but rather to a mental space, abstract in nature. Of course, the mental space of human concepts, in the typical view of materialism and reductionism, originates precisely in the activity of the human brains, which are naturally contained in the ordinary spatial theater. But independently of the fact that human concepts originate from (or only from) our specific localized brain-structures, the fact remains that the spatiality of a human concept, such as for example the concept ``fruit,'' is quite different from the spatiality of an ordinary physical object.

In fact, the concept ``fruit'' being an abstract entity, it doesn't actually possess, in a stable way, the property of \emph{localization-1}, which is instead typical of concrete objects, while it unquestionably possesses the property of \emph{localization-2}, since the fruit-concept is always fully available in interacting with specific semantic entities, for instance formed by specific \emph{aggregates of concepts}, in the context of specific phrases, to produce a predetermined localization.

In other words, in the same way an electron is able to relate/bind (in an ephemeral or stable way, depending on the type of experimental context) to a specific macroscopic system, thus manifesting its presence in the three-dimensional theater, also the human concept ``fruit'' can temporarily assume the status of an object, when it relates/binds with a specific \emph{objectifying} context, for instance in the ambit of the following injunctive sentence: ``Look at the fruit which in this moment is on the table!''

If on the table in question there are two fruits, an apple and a pear, the experimental context is such that in order to act the above-mentioned injunction a human being must \emph{choose} which of the two object-fruits to stare at, by selecting a specific \emph{visual interaction}, attributing in this way to the abstract concept ``fruit'' a specific ephemereal objectual form, and the corresponding spatial localization associated to it. And since the decision process is usually quite sensitive to \emph{intrapsychic} and \emph{extrapsychic} fluctuations, in no way it will be in general predictable in advance. Therefore, we are in a situation which is akin to the observation of the localization-1 property of a microscopic entity: as is the case for an electron, the abstract human concept ``fruit'' doesn't ordinarily possess, in actual terms, the localization-1 property.\\
\begin{figure}[!ht]
\centering
\includegraphics[scale =1]{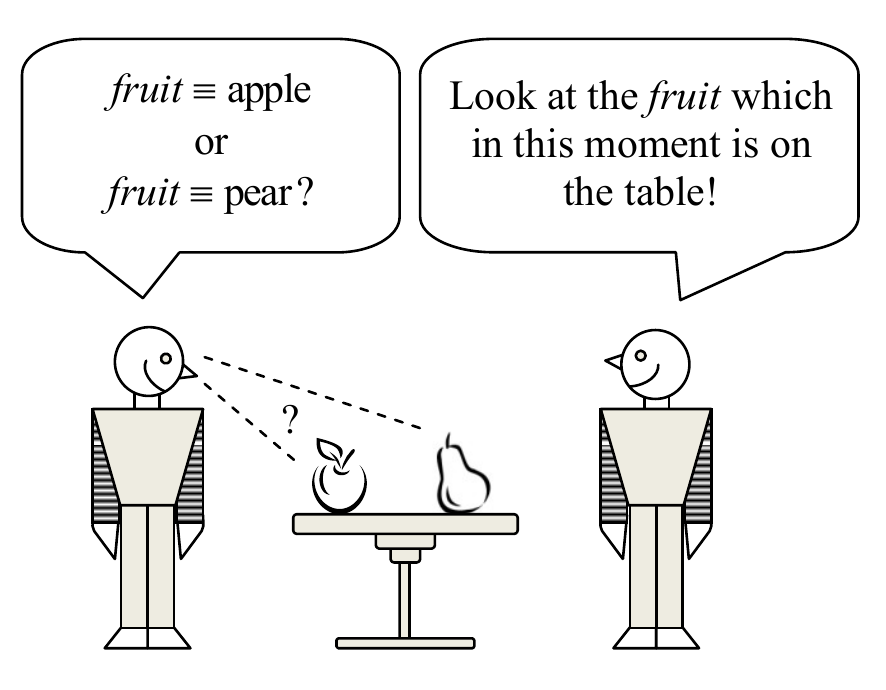}
\caption{The ``fruit'' abstract concept is ephemerally objectified into the concrete concept of the apple or the pear, in a way which is generally unpredictable, by means of the experimental context expressed by the injunctive phrase ``Look at the fruit which in this moment is on the table!''
\label{melapera1}}
\end{figure}

On the other hand, if we modify the experimental context, replacing the injunctive sentence by the following one: ``Look at the fruit which in this moment is \emph{closer to you}, on the table!'' then the human being acting the injunction will not anymore relate to the pear, and this means that this new context is now \emph{forcing} the concept ``fruit'' to identify with a perfectly predeterminable object, and assume its spatial localization, which is the one of the apple. Thus, we are in a situation corresponding to the observation of the classical localization-2 property.\\
\begin{figure}[!ht]
\centering
\includegraphics[scale =1]{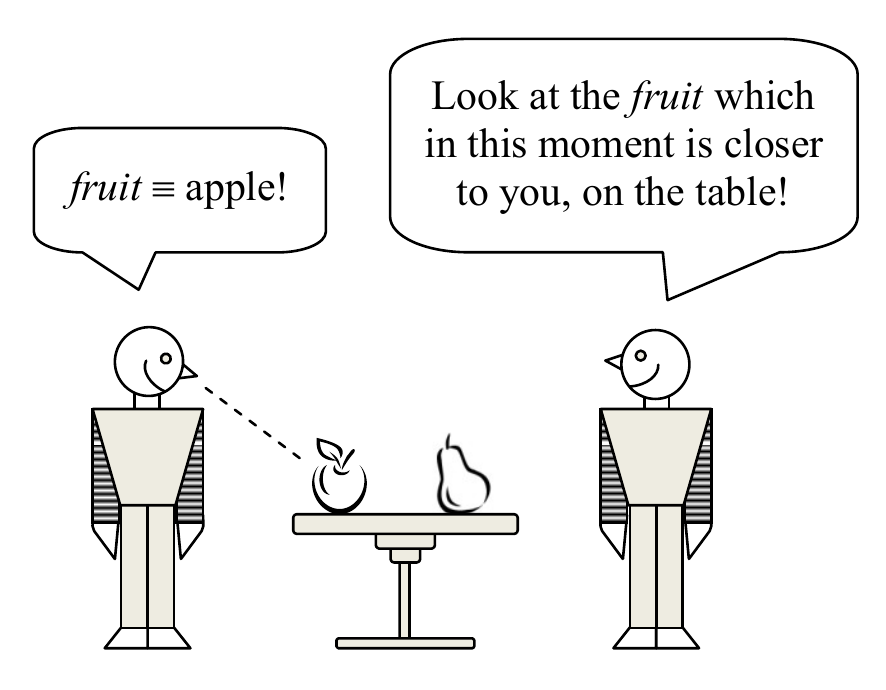}
\caption{The ``fruit'' abstract concept is stably objectified in the concrete concept of the apple present on the left side of the table,
by means of the determinative experimental context expressed by the injunctive phrase ``Look at the fruit which in this moment is closer to you, on the table!'' (The additional fragment of sentence ``closer to you'' plays here the same role as the repulsive field in the case of the electron, or the scissor in the case of the elastic band).
\label{melapera2}}
\end{figure}

As this simple example illustrates, the non-spatiality of human abstract concepts is very similar to the non-spatiality of quantum microscopic entities. It is important however to emphasize that the analogy between human concepts and microscopic entities, considered as conceptual entities, is much deeper and articulated than what could be understood by the abovementioned fruit-example, as demonstrated by the careful analysis of Aerts and collaborators. In fact, human concepts are able to express almost all the complex phenomena typical of the microscopic quantum level, such as \emph{entanglement}, violation of \emph{Bell's inequalities}, \emph{superposition} and \emph{interferences}, \emph{coherence} (which in the ambit of human concepts becomes connection through meaning), incompatibility (expression of the impossibility for a concept of being simultaneously maximally abstract and concrete), etc.~\cite{Aerts-c1,Aerts-c2,Aerts-c3,Aerts-c4}. 

But while taking into account these similarities, it is important not to fall victims of too easy anthropomorphisms and confuse human concepts with the microscopic quantum entities, or human minds with the measuring apparatus in a laboratory. What the profound analysis of Aerts teaches us is simply that human concepts are not the only conceptual entities we humans are dealing with, since also the entities of the microscopic world, like electrons, have that typical behavior that we humans usually attach to concepts, and not to objects.

\section{Control versus discrimination}
\label{Lack of control}

Our parallel between the lateralization property of an elastic band, the objectification process of an abstract human concept, like a fruit, and the localization property of a microscopic entity, like an electron, offers us a further argument in support of our thesis about the presence of an irreducible form of indeterminism in our description of reality. Indeed, if we really want to deterministically observe an electron in a given region of space, nothing in principle impedes us to do so: we only have to force it to manifest in that region, by precluding its manifestation in any other place. This kind of observational procedure, which is a creation-2 process, cannot however be considered as a spatial detection of the electron in the ordinary sense. Indeed, detection doesn't mean \emph{forced capture}, i.e., to coerce the electron to appear in a specific place, but just to observe if it will appear in a specific place, when it is allowed to do so (for instance because a detection instrument is placed in that region).

This doesn't mean though that the observational process will be for this non-invasive: interacting with a detection instrument (like for instance a screen) is undoubtedly an invasive process, which will sensibly alter the state of the electron. But we must distinguish here two different levels of invasiveness: there is the level which may be intrinsically incorporated in the observational process itself, and therefore cannot be eliminated (like when we observe the burnability of a piece of wood), and there is the level which depends on the amount of control we want to exert in the experimental procedure, to make it more or less deterministic. A perfectly controlled observational experiment is a procedure such that the entity under investigation is actively acted upon in a way as to limit its degrees of freedom, so that only a specific single outcome will be left open, and therefore will be necessarily actualized.  

Although experimenting with entities by means of perfectly controlled, deterministic observational processes can certainly reveal many of their important characteristics (especially when the observational processes are of the discovery kind), it cannot reveal all of them. For instance, considering the fully controlled observation of the localization-2 property, it is clear that an electron and a macroscopic body both possess such a property in actuality. But this means that such a controlled observation, where the effects of fluctuations in the selection of outcomes is filtered out, will not be able to reveal the deep difference between the non-locality (i.e., non-spatiality) of a microscopic entity, like an electron, and the typical locality (spatiality) of a macroscopic corpuscle. 

In other terms, by only relying on fully controlled procedures, we cannot fully characterize the different entities forming our reality and form a complete picture of them. It is in fact in the ambit of uncontrolled observations of localization-1 properties that microscopic entities like electrons are able to reveal, among other things, their genuine non-ordinary spatial nature (i.e., their locality of type 1: the fact that in general they can only enter the three-dimensional physical space in an ephemeral way), that distinguishes them from macroscopic local bodies, which belong to the three-dimensional space in a stable way.

The same is true, \emph{mutatis mutandis}, for elastic bands. Indeed, imagine that we don't know exactly how the elastics have been manufactured. For instance, let us assume that although exteriorly the elastics look uniform, in the sense that their longitudinal section is constant all along their length, interiorly  they are made of a non-uniform material, in the sense that there are specific segments of the elastic bands which are more easy to break than others, when the elastics are stretched (for instance because the density of the rubber is not uniform)~\footnote{Clearly, for non-uniform elastic bands, the ``scissor-RNG'' procedure (Def.~\ref{definition-6}) cannot anymore be considered as equivalent to the original procedure (Def.~\ref{definition-5}).}. Then, the structural difference between uniform and non-uniform elastic bands cannot be discriminated by only performing fully controlled observations of the left-handedness-2 property, as such property is always actual for both uniform and non-uniform elastics. On the other hand, their difference may be revealed by analyzing the statistics of the outcomes of the indeterministic observation of left-handedness-1 property (provided of course that the non-uniform elastics are all identical, the observations are performed a sufficient number of times and the bands are always grabbed with the same orientation).~\footnote{The difference in the statistics of outcomes when using non-uniform elastic bands, instead of uniform ones, has been extensively analyzed in the so-called $\epsilon$-model. In this ambit, it can be shown that non-uniform elastics can give rise to quantum-like, intermediate behaviors, neither purely classical nor purely quantum~\cite{Aerts3}.}

The moral of the story is that although an increase of experimental control allows for more predictable outcomes, at the same time it severely limits our ability to bring about all the structural differences among the different entities populating our reality. In some sense, the greater is our control in determining specific outcomes, the lower will be our discriminative power, i.e., our ability to grasp the full spectrum of characteristics of the entity under investigation. This is so also because, as we tried to emphasize in this paper, increasing the control over the observation of a specific property will inevitably radically alter the observational process, so that the property which is observed will not any more be the same property.

Thus, although we certainly agree with Aerts when he states that~\cite{Aerts4} ``There is [...] no problem with the deterministic hypothesis. There is no incompatibility at all between quantum mechanics and a complete deterministic world as a whole, since the probabilities appearing in the quantum theory can be explained as being due to a lack of knowledge about the interaction between the measuring apparatus and the entity during the measurement, and hence are of epistemic nature,'' we think such a statement should be understood in the sense that: \emph{there is no incompatibility between quantum mechanics and determinism, as all interactions between physical entities and measuring apparatus can be assumed to be deterministic; however, this doesn't allow us to conclude that all observational processes are also deterministic, or could be made such, as genuinely indeterministic creation-1 processes are a key ingredient in our discriminative analysis of the physical reality, and we cannot get rid of them if we want to obtain the most possible complete picture of it}.

\section{Relational properties}
\label{Relational properties}

In the previous section we have emphasized that our reality is richer than what purely deterministic, fully controlled observations would allow us to establish: if we want a complete picture of reality, then we need to also include in our observations a condition of lack of control, from which indeterminism, in the sense of a genuine unpredictability, results. In this section, we want to consider how indeterminism is responsible for the emergence of \emph{relational properties} in typical quantum observational processes, and therefore to their observed ephemerality, as we already emphasized in Section~\ref{Quantum elastic bands}.

The concept of \emph{relational properties}, as opposed to the concept of \emph{intrinsic properties}, was discussed by the present author in a recent paper~\cite{Massimiliano-O}, where the difference between intrinsic properties, that an entity can possess independently of a specific observer (and therefore are equivalently observable by different observers), and (relational) properties, characterized by a specific configuration of the composite system formed by the observed entity plus a specific observation instrument, was emphasized. 

Before stating our definition of relational property, we would like to mention that the relevance of relational properties in quantum mechanics was also discussed some time ago by Gomatam~\cite{Gomatam}. When we analyzed relational properties in~\cite{Massimiliano2}, we weren't aware of~\cite{Gomatam}, and therefore we take the opportunity of the present discussion to put our analysis also in the perspective of what has been proposed by Gomatam.

Gomatam rightly pointed out that:~\cite{Gomatam} ``[...] in a consistent interpretation of quantum theory, if the macroscopic world would also have a quantum description, then a macroscopic object must possess more physical properties than those accounted for by classical mechanics.'' This is what we have also emphasized in Section~\ref{Third objection} of this paper, when we have distinguished ordinary properties from non-ordinary ones. 

Gomatam pursues his discussion proposing that relational properties could correspond to these additional properties a classical object can have. He distinguishes them from what he calls \emph{primary properties} (i.e., intrinsic properties, in the language of the present article), by means of the following characterization:~\cite{Gomatam} ``[...] while the primary properties are only \emph{expressed by} a relation with another object (such as a scale or a clock), the relational properties are expressed by and \emph{actualized} in a relation with another object.''

In other terms, according to Gomatam, the characteristic of a relational property, as opposed to an intrinsic one, is that a relational property would be an actual property only in the moment it is an observed property. As an example, Gomatam considers a book, which clearly possesses intrinsic (primary) properties (like having a certain mass, volume, etc), but can also be used, for instance, as a \emph{paperweight}. He then suggests that \emph{paperweightness} would precisely be a relational property that, I quote:~\cite{Gomatam} ``can be regarded as a potential property of the object, which becomes `physically real' only when the object is placed in an appropriate spatio-temporal relation with another object'' (like a stack of paper). 

What Gomatam however doesn't explain is why a relational property like the paperweightness of a book would only be actualized when the book exhibits a specific relation with the stack of paper-measuring apparatus. Why such a property cannot be possessed by the book also when it doesn't entertain such specific relation with a stack of paper? To answer this fundamental question, we need to complete Gomatam's example of paperweightness by providing a specific operational definition of such a property. Only in this way it becomes possible to truly understand the essence of what a relational property is, and provide what we think is a general definition.  

An important point to emphasize is that paperweightness can conventionally be defined in different, non equivalent ways, and according to the chosen definition it can either be understood as an intrinsic property of a book, or as a relational one. More precisely, let us here distinguish \emph{paperweightness-1} from \emph{paperweightness-2}, as we distinguished in the previous sections left-handedness-1 from left-handedness-2, and localization-1 from localization-2.

\begin{mydef}
\label{definition-9}
\emph{[paperweightness-1 of a book]}. In the location where the book is present (say, an office) place a suitable number of stacks of paper and of human beings~\footnote{What a \emph{suitable} number exactly means should of course be specified, but this would unnecessarily complicate the discussion. Let us simply assume that we possess a criterion, which we don't need to specify here, to determine the exact number of stacks of paper and human beings to be used in a given location, considering for instance its dimensions}. Then monitor the location for one hour, by means of a hidden camera. If, before the end of that time period, the persons in that location place the book on one of the stacks of paper, then paperweightness-1 has been successfully observed; otherwise, the observation is to be considered unsuccessful.
\end{mydef}

\begin{mydef}
\label{definition-10}
\emph{[paperweightness-2 of a book]}. Take the book and place it on a stack of paper of your choice. If you succeed in performing this action, then paperweightness-2 has been successfully observed; otherwise, the observation is to be considered unsuccessful.
\end{mydef}

It is important to understand the difference between these two properties: paperweightness-1 is a relational property of the book, whereas paperweightness-2 is an intrinsic property of the book. That paperweightness-2 is an actual, intrinsic property of the book, is clear from the fact that the observational process that defines it is of the creation-2 kind: it changes the position-state of the book in a deterministic way, creating a specific relation between the book and a given stack of paper. A book always possesses such a property, even when it is not placed on a stack of paper, as we can predict with certainty that, should we decide to observe it, the observation would be certainly successful. 

What about the paperweightness-1 property of a book? In this case it is not anymore about observing the availability of the book in lending itself to the action of a human being, in order to be positioned in a specific spatial relation with a stack of paper: paperweightness-1 is about observing the effective availability of the book in manifesting such a relation, when a certain number of stacks of paper and human beings are present in its environment, that we haven't instructed in whatsoever way to act deterministically in order to create such a relation. 

This means that the definition of paperweightness-1 do not contemplate the possibility for the observer to predict in advance, with certainty, the outcome of the observation. When s/he looks to the book during the one hour time period required by protocol, s/he may see it or not on a stack of paper, and this depends on the way the humans who are present in the environment will think about using the book. They may leave it in its initial position, place it on the shelves of a library, read it, or maybe use it as a paperweight. This depends on a number of fluctuating factors which are not under the control of the observer, considering how paperweightness-1 has been defined. 

A question then arises: is the observation of paperweightness-1 a creation-1 process or a discovery-1 process (see Def.~\ref{definition-1} and Def.~\ref{definition-2})? Actually, considering the way we have defined it, it's clearly a creation-1 process. Indeed, as a result of the observation, the book state will possibly change (by changing its spatial position and orientation) and this change of state is provoked by the fact that we have placed in the location under consideration a certain number of stacks of papers and of human beings which, together with the hidden camera, are part of the observer's measuring apparatus. 

On the other hand, if we consider that the stacks of paper and human beings are part of the natural environment of the book-entity, then the observer with her/his hidden camera is not going to alter, in any way, the evolution of the book-entity, so that the observation could now to be considered a discovery-1 process. In principle, every creation-1 process can be reframed as a \emph{discovery-1} process, if we \emph{conventionally} describe the measuring apparatus as part of the entity's natural environment. However, if we do this, we must keep in mind that we are not anymore observing the same property, nor actually the same system (see the discussion of Sec.~\ref{Third objection}). 

The reader will have certainly noticed the similarity between the paperweightness-1 property of a book and the localization-1 property of a microscopic particle. The stacks of paper play here the same role for the book than the position detectors for a microscopic entity. When a microscopic entity, say an electron, is not present in space, it is like a book which is not placed on a stack of paper; and when the electron is detected in space, i.e., when it creates, momentarily or stably, a relation with a specific macroscopic entity, it is like the book establishing a relation with a stack of paper, being positioned on top of it. Having said this, let us now conclude our analysis and explain why paperweightness-1 is genuinely relational, whereas paperweightness-2 is not. For this, it is time to provide our specific definition of what a relational property is: 

\begin{mydef}
\label{definition-11}
\emph{[relational property]}. A property associated to an entity is \emph{relational} if and only if there are states of the entity for which the observations of the property by different observers are not \emph{deterministically equivalent}.
\end{mydef}

The term ``observer'' in the above definition is to be understood in the sense of an \emph{instrument of observation} (plus possibly a human being able to interpret the outcomes of the process, according to a given protocol). By the term ``deterministically equivalent,'' we mean equivalency as per point~(1) of Def.~\ref{definition-4}.
	
Obviously, Def.~\ref{definition-11} hints to the fact that a relational property is about an \emph{exclusive} relation between the system under consideration and a specific observer. In~\cite{Massimiliano-O}, as a paradigmatic example, we have considered the property of a macroscopic body of having its center of mass in a specific spatial position $\textbf{x}$. The property is relational because a different observer, associated to different reference frame, will necessarily observe, at the same time, a different values $ \textbf{x'}\neq\textbf{x}$ for the body's center of mass position, so that their observations cannot be considered to be deterministically equivalent. To have a specific position in space is in fact to have a specific relation with a reference frame which makes such a position manifest. 
	
Consider now the observation of properties like the left-handedness-1 of an elastic band, the $R$-localization-1 of an electron, or the paperweightness-1 of a book. These are also relational properties, but for a different reason than the previous example. Here the relational aspect of the property is due to the fact that the outcome of an observational process is genuinely unpredictable. Thus, observations by different observers are not \emph{deterministically equivalent}, but only \emph{indeterministically equivalent}, in the sense specified by point~(2) of Def.~\ref{definition-4}.

In other terms, it is the genuine unpredictability of the outcome of the observation that makes the above mentioned ``type 1'' properties the expression of an \emph{exclusive} relation of the entity with a specific observer (i.e., with a specific measuring apparatus). Indeed, if a same observation would have been performed by a different observer, at the same moment, a different outcome would have possibly resulted.

This of course is also true for the paperweightness-1 property, since a ``different observer'' means here a different way to implement the observational protocol, for instance by placing the stacks of paper in a slightly different way in the location, or choosing different persons to participate in the experiment, as these differences, however small they are, certainly have the capacity to affect the final outcome.

However, if we would have decided to define differently the paperweightness-1 property, as a purely non-invasive observational process, considering for this that the stacks of papers and the persons present in the location would be part of the natural environment of the book-entity, then of course the observational protocol would simply be about the observation, by means of a hidden camera, of a possible relation of the book-entity with some others entities present in the environment. But this observation can be equivalently performed by different observers, and therefore cannot anymore be considered as a relational property of the book-entity. However, as we emphasized many times in this article, a paperweightness-1 redefined in this way would not correspond any more to the same property.

The analysis of this section allows us to clarify our statement about quantum measurements which, as we briefly discussed in Sec.~\ref{Quantum elastic bands}, would result from~\cite{Massimiliano-O}: (1) the presence of a random selection mechanism; (2) of an irreducible invasiveness of the observational process; and (3) of the ephemerality of the observed property, due to its relational character. We now understand that point~(3) is in fact not independent from points (1) and (2), but actually a direct consequence of them. Indeed, the invasiveness of the process is responsible for the \emph{creation} of a specific outcome, i.e, of a specific relation between the observed entity and the measuring apparatus. But because of the random mechanism that selects/creates the outcome, such a relation is unique to that specific measuring apparatus, and to no other. Therefore, it is not the expression of an intrinsic property, of the entity under consideration, put of a relational one, according to Def.~\ref{definition-11}.

This relational character, in turn, is responsible for the typical ephemerality of quantum properties~\cite{Massimiliano1,Massimiliano-O}. Indeed, once an observational process is completed, usually the exclusive relation between the entity and the measuring apparatus is broken (in the case of lefthandedness-1, this happens when the observer's hands let go the elastic fragments), so that to observe the property in question again, one has to recreate it, by following once more the corresponding non-deterministic observational procedure, which of course doesn't guarantee that the outcome will be again successful. Hence the ephemerality of the property, which can only be attributed to the entity in the very moment it is observed.~\footnote{In~\cite{Massimiliano-O} we have remarked that the positions measured in Aert's quantum machine~\cite{Aerts3,Aerts4,Aerts13} are not relational properties. This may appears in contradiction with our present analysis, considering that positions are observed by means of observational processes that contain a random selection mechanism and an irreducible invasiveness. However, a closer analysis reveals that the surface of the three-dimensional Euclidean sphere on which the particle lives, plays in fact a double role, as it is also part of the measuring apparatus. Therefore, one cannot in this model completely separate the measuring apparatus from the entity, which is the reason why at the end of the observation the property remain actual and can be re-observed with certainty.}

We conclude this section with a last remark. One should not confuse the present concept of relationality with the one of Rovelli~\cite{Rovelli}. In so-called \emph{relational quantum mechanics}, introduced by Rovelli, it is the state itself of a physical entity which is considered to be observer-dependent. On the contrary, in the present paradigm, the state of a physical entity is an observer-independent, intrinsic attribute to the entity under consideration, whereas the relational character of certain properties emerges as a consequence of the specificities of the observational protocols that are used to observe them.

\section{Conclusion}
\label{Conclusion}

In this article we have analyzed how indeterminism intervenes in some of our observational processes, as an integral part of them. We will not recapitulate in this last section our findings, but offer instead a conclusive thought about the relation of our analysis with the old debate about free choice.    

First of all, we observe that the concept of \emph{free choice} and the one of \emph{free will}, although usually confused, are not the same. Free will, as its name indicates, mostly refers to the hypothetical ability of humans (and possibly of some animals) to produce a ``volitional causation,'' that would not be explainable in terms only of deterministic physical laws. The existence of a genuine human free will, understood in this sense, is therefore beyond the scope of physics, as usually understood, and the description of its effects in the physical world would in fact constitute a sort of extra-physical addition to the presently known physical laws. 

This is the reason why, although people usually believe in the existence of a genuine volitional causation, as they subjectively experience it, because of the metaphysical implications of such a notion many scientists consider it to be a mere illusion, according to the view that all physical entities, humans or not humans, would necessarily be subjected to deterministic (physical) laws. 

But apart from the metaphysical problem of a human free will independent of physical deterministic causation, there is the simpler question of \emph{free choice}. As its name indicates, free choice is not about ``will,'' but about ``choice,'' and choice, in turn, refers to the existence of \emph{possibilities}. If there is only a single possibility in a process, then there is no choice, but if more possibilities are present, then there is real choice. However, if the mechanism selecting a possibility is in principle predictable by the one who selects it, obviously there would only be illusion of possibilities, and therefore only illusion of choice. 

Now, as we emphasized in the final paragraph of Sec.~\ref{Lack of control}, there is no incompatibility between the existence of deterministic laws governing our reality and the emergence of a genuine form of indeterminism in our observations. When we consider our world in global terms, as a sort of ``block entity,'' we can certainly affirm that, in a sense, everything is as it should be, according to the deterministic laws operating at all levels of reality. But a global view is different from a local one, where the term ``local'' is  here not to be  understood only in the reduced sense of spatial locality. 

Indeed, we can easily consider that choice would be non-existent at the global level, but nevertheless truly existent at a local one, in our reality. This because when we ask an operational question, and associate to it (consciously or unconsciously) a practical procedure to obtain an answer, then according to the specificities of the procedure (that could also be intrapsychic) we may not be in a position to predict such answer in advance, not even in principle. And if the answer is what determines our choice, in a given circumstance, then being our choice truly unpredictable, in that sense (but only in that sense) it can be considered a non-illusionary free choice.

\end{document}